# Efficient Analysis and Visualization of High-Resolution Computed Tomography Data for the Exploration of Enclosed Cuneiform Tablets

Stephan Olbrich,
Andreas Beckert*

Universität Hamburg

Cécile Michel

Centre National de la Recherche Scientifique
(CNRS – ArScAn), Nanterre, and Universität Hamburg

Christian Schroer, Samaneh Ehteram,
Andreas Schropp, Philipp Paetzold

Deutsches Elektronen-Synchrotron (DESY),
Hamburg, and Universität Hamburg

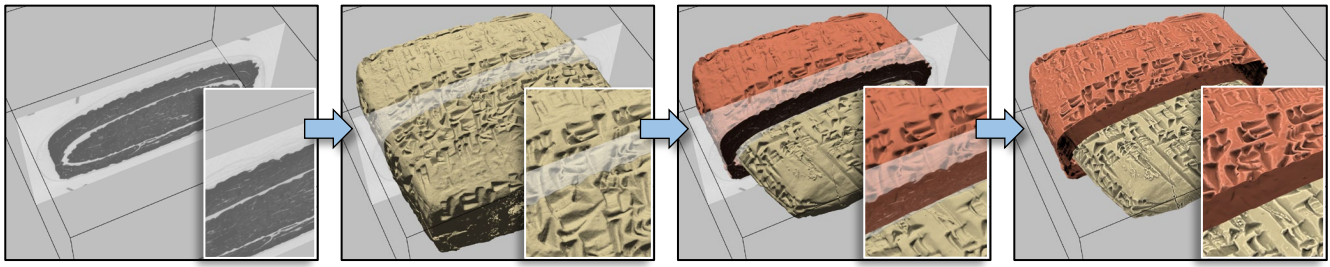

Figure 1: Steps of processing density data on a regular 3D grid, acquired by high-resolution computed tomography of an enclosed cuneiform clay tablet: Volume data denoising, surface reconstruction, smoothing, segmentation, and feature-enhanced rendering.

## ABSTRACT

Cuneiform is the earliest known system of writing, first developed for the Sumerian language of southern Mesopotamia in the second half of the 4$^{th}$ millennium BC. Cuneiform signs are obtained by impressing a stylus on fresh clay tablets. For certain purposes, e.g. authentication by seal imprint, some cuneiform tablets were enclosed in clay envelopes, which cannot be opened without destroying them. The aim of our interdisciplinary project is the non-invasive study of clay tablets. A portable X-ray micro-CT scanner is developed to acquire density data of such artifacts on a high-resolution, regular 3D grid at collection sites. The resulting volume data is processed through feature-preserving denoising, extraction of high-accuracy surfaces using a manifold dual marching cubes algorithm and extraction of local features by enhanced curvature rendering and ambient occlusion. For the non-invasive study of cuneiform inscriptions, the tablet is virtually separated from its envelope by curvature-based segmentation. The computational- and data-intensive algorithms are optimized for near-real-time offline usage with limited resources at collection sites. To visualize the complexity-reduced and octree-based compressed representation of surfaces, we develop and implement an interactive application. To facilitate the analysis of such clay tablets, we implement shape-based feature extraction algorithms to enhance cuneiform recognition. Our workflow supports innovative 3D display and interaction techniques such as autostereoscopic displays and gesture control.

## 1 INTRODUCTION

Cuneiform is the earliest known system of writing. This word means wedge-shaped and describes the appearance of the signs, which combine various wedges obtained by impressing a stylus into the fresh clay of a tablet. It was first developed for the Sumerian language in Southern Iraq during the second half of the 4$^{th}$ millennium BC and was in use during more than three millennia for a dozen different languages. Cuneiform signs are written from left to right, and all the sides of the clay tablet are successively covered by rotating it along a horizontal axis [31]. Today, hundreds of thousands of cuneiform texts have been deciphered and studied by Assyriologists in order to write the history of the ancient Middle East.

---

* {stephan.olbrich, andreas.beckert}@uni-hamburg.de

Clay cuneiform tablets written for practical purposes, such as letters or legal documents, were sometimes enclosed in clay envelopes [5]. The purpose of the envelope was to protect the confidentiality of the text and the material integrity of the tablet. The envelope of the legal text certified the validity of the document [30]. For legal documents, a summary or the transaction was copied on the envelope, and parties and witnesses unrolled their cylinder seals on the surface of the clay leaving miniature scenes with the value of signatures. When the envelope of a contract was broken, it lost its legal validity. Hundreds of such legal tablets have been discovered, still encased in their clay envelope, and thus the content remains invisible to Assyriologists. However, there may be important information on the tablet that has not been included on the envelope.

These enveloped cuneiform tablets are interesting objects of research in Assyriology, but could so far only be studied if the envelope would be broken (Figure 2).

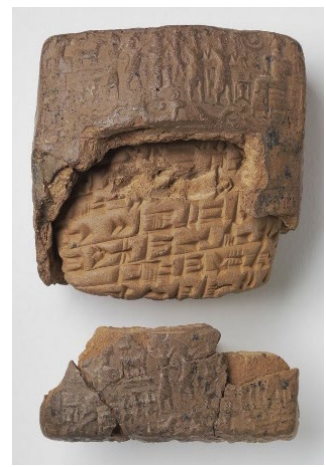

Figure 2: Cuneiform tablet and envelope, Old Assyrian, 1927-1836 B.C. (www.harvardartmuseums.org).

The aim of our project is to extract cuneiform text non-destructively from enclosed tablets and to reconstruct how they were made. The workflow to be designed must be applied at the collection site, as it is not permitted to take these ancient, unique artifacts outside. This is the case for the Louvre Museum (Paris), where one of the world-wide most relevant collections of cuneiform tablets exists.

To study enclosed tablets non-destructively, the idea is to scan the tablets using computed tomography (CT) and to reconstruct the artifacts on a high-resolution regular 3D grid. X-ray micro-CT instruments exist, but they cannot easily be brought to a collection (e.g. in a museum), due to their size and weight. For these reasons, as part of our interdisciplinary project, a portable high-energy X-ray CT system has been developed. The system must be completely self-contained, with control and data processing working offline. This comprises compute resources, data capacities, visualization facilities, and its efficient exploitation on location, where no internet connection can be expected. The CT scanner weighs about 420 kg. For transport, it can be disassembled into several parts, each of which can be carried by two or three people if necessary. The system can be assembled and operational in less than two hours (Figure 3).

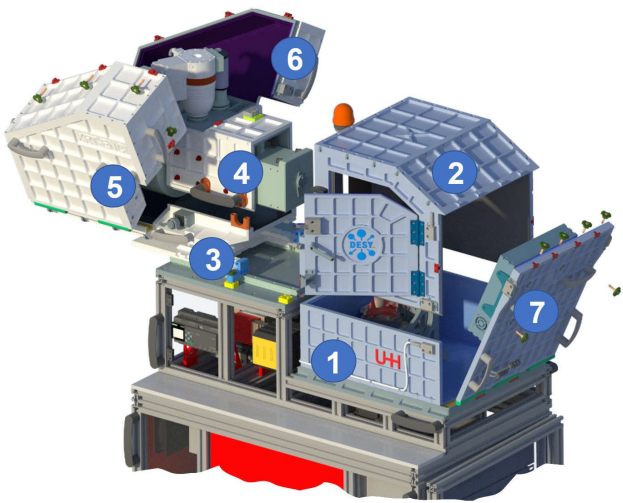

Figure 3: Individual modules of the ENCI X-ray system: (1) Support frame with lower radiation box, (2) Upper radiation box, (3) X-ray tube bottom, (4) X-ray tube, (5) X-ray tube shielding, right side, (6) X-ray tube shielding, left side, (7) Back-side wall, detector.

The micro-CT scanner is named ENCI (Extracting Non-destructively Cuneiform Inscriptions), in homage to ENKI, the Sumerian god of water, knowledge, crafts, and creation. It fulfills the requirements of high resolution to reproduce details of inscriptions (microfocus X-ray source size: 20 μm, detector: 3072 x 1944 pixel, 74.8 μm pixel size), supports maximum sample dimensions of 10 cm x 20 cm x 5 cm, providing sufficient X-ray energy to penetrate clay material with these dimensions ($E_{max} = 180$ keV) in reasonable tomographic recording time (26 frames per second, multiplied by thousands of projections, resulting in measuring time of approx. 20 minutes), and appropriate radiation protection shielding to overcome international certification requirements (< 0.5 μSv/h outside of the instrument).

The measurement of the artifacts by ENCI results in a volume representation, reconstructed on a regular 3D grid, provided in HDF5, a scientific open file format, which is a convenient basis for data analysis and visualization afterwards. The spatial resolution varies, depending on the object, and the disposition in the instrument. The dataset represents up to 3072 x 3072 x 1944 grid points (approx. 18 billion voxels), where scalar values in 32-bit float precision are stored (approx. 72 GB).

However, the desired application scenarios, such as virtually unpacking the enclosed cuneiform tablet and reading its inscriptions, are not possible with direct visualization methods for volume data, such as volume slicing (first step in Figure 1) or volume rendering. Volume rendering requires sophisticated preprocessing and usually manual efforts to support the segmentation of the tablet and the envelope. Also, 3D surface reconstructions usually require time-consuming, explorative manual operation. This approach could be useful for material analysis to study the properties of the artifacts. However, the method we present here includes automatic geometric surface extraction and separation of the tablet and the envelope, as well as 3D rendering and 3D printing. Tight schedules for data acquisition at the collection sites, directly followed by interactive exploration of the artifacts, require time-efficient data processing and rendering.

For these reasons, we set up the following requirements for the processing of the acquired micro-CT data:

1. Automated parameterization of the compute- and data-intensive processing algorithms, to avoid high-latency time-consuming manual operation, as far as possible
2. Efficient implementation for optimized workstation utilization on location, esp. exploiting multi-level parallelization
3. Decomposition of high-resolution volume data preprocessing (batch-oriented), and interactive visualization of extracts
4. Development and implementation of tools for interactive 3D exploration of extracted, segmented, and attributed surfaces, optionally supporting immersive 3D displays and interaction
5. Design of a compressed intermediate representation of surfaces, to support efficient transfer of data from the preprocessing tool to the visualization application
6. Triangulated surfaces should satisfy topologically 2-manifold criteria, to support universal mesh reuse
7. Support of open file formats, such as HDF5 for 3D micro-CT data import, STL for 3D printing, TIFF export of screenshots, image sequences of keyframe animations, and fat-cross renderings, and export of OBJ and PLY files for photorealistic rendering by using separate applications (e.g. Blender)

## 2  RELATED WORK

Specialized tools to analyze and visualize cuneiform tablets, have been developed for more than two decades. First studies were based on surface representations, i.e. by techniques such as laser-based Light Detection and Ranging (LiDAR), Structure from Motion (SfM) or Structured Light Scanning (SLS) [1][8][15]. However, such approaches only digitize the exterior of the tablets and do not allow the study of tablets and their inscriptions enclosed in a clay envelope, since digitalization of the interiour would require breaking the artifact.

First experiments of X-ray CT scanning to explore the interiour of cuneiform tablets were performed at Delft University of Technology (Netherlands), where ancient artifacts of the Liagre Böhl Collection of The Netherland Institute for the Near East (NINO, Leiden, Netherlands) have been scanned and visualized [36][55]. Reconstructed triangulated surfaces have been analyzed and rendered, which already enabled to read cuneiform signs of the artifact and explore features, such as plant materials and holes in the clay [32]. In [16], it was shown that these meshes can be imported in GigaMesh [15], a software framework with focussed on processing surface meshes as described above, and segmented by interactively applying algorithms which result in virtually unpacking a cuneiform tablet. However, this approach, as well as generic tools (Avizo [3], VTK [48]), commonly applied for interactive visual data analysis, lacks automatic extraction, segmentation, and interactive visualization of the separate parts of the artifacts, its inscriptions and other features contained in the clay tablet.

Transfering methods from 2D image segmentation to 3D CT data can be useful [43]. Alternatively, features in the 3D images, such as ambient occlusion, could be exploited for classification and separation [53]. This approach results in enumerated partition information per voxel, thus does not provide hints for subpixel precision, and it is computational- and data-intensive. We decided to concentrate on segmentation of meshes. It has been shown that automatic mesh segmentation requires exploitation of application-specific characteristics, e.g. shape features [49].

To reconstruct the surfaces between clay material, surrounding air, or other material inside, with less density than clay (air, organic matter etc.), isosurface extraction methods can be applied. The well-known marching cubes algorithm [25] creates a "triangle soup", and mesh structures have to be constructed in a subsequent step. Ambiguous configurations can result in discontinuities or violation of the 2-manifold condition, which is presupposed for mesh processing, such as smoothing operations, and for 3D print. Dual marching cubes algorithms [11][46] are advantageous regarding resulting triangle quality and implicit neighborhood information. Since they do not neccessarily fulfill the manifold topology, the algorithms have been extended appropriately [47]. Ambiguities have been resolved, first by analyzing the cube faces [35], and later based on volumetric models and analyzing topological aspects [6][7][9][21][24][34]. This resulted in an increased number of finally 37 unique topological cases [6]. Approaches for compressed and multiresolution of isosurfaces has been presented, e.g. [12][22][43][45][52]. The best known compression ratio is achieved in [19], but it finally depends on context-based arithmetic coding, which is executed sequentially, i.e. parallelization of it would be nontrivial.

Denoising raw 3D CT data is required to achieve good results from contouring (Figure 4). For this purpose, median and bilateral image filtering algorithms are commonly used [10][42][54]. In joint bilateral filtering, the range weights are calculated from other image data than the original image to be filtered. Instead of the original gaussian range weighting function, alternatives have been suggested, that can be more effective regarding denoising quality, at the same time significantly reducing the computational effort, e.g. the Tukey weight function [10]. The fundamental concepts are reused for feature-preserving smoothing of surface meshes. In comparison to first approaches of filtering vertex positions [14], processing of face normals in a first step, and adapting vertex positions afterwards, is more effective, and usually performed iteratively [20][26][51][59]. For median filtering, instead of replicating normal vectors for higher weights, and sorting them (as in [57][58]), an iterative method based on [50] can be used. The evaluation regarding quality and performance, as well as further improvements of this approach are addressed in our work.

For better recognition of inscriptions, conventional lighting and material models are complemented by non-photorealistic rendering of triangulated surfaces [13][41] and effective shadowing, e.g. by applying the method of ambient occlusion [28][29].

The algorithms implemented in available applications, lack in consequent integration of multi-level parallelism in all processing stages, in considering all special cases of data configurations for 2-manifold marching cubes, in exploitation of implicitly existing information by combining operations instead of reconstruction afterwards, and in chosing most effective algorithms, resulting in a tradeoff, regarding run time and quality of the results.

To overcome the insufficiencies of existing algorithms and applications, we developed an integrated workflow, implemented and customized for efficient support of the cuneiform exploration scenario, and meeting the requirements stated at the end of our introduction (Chapter 1). First results, based on CT data of a cuneiform replica by the use of a preliminary CT scanner, were presented in [36][37].

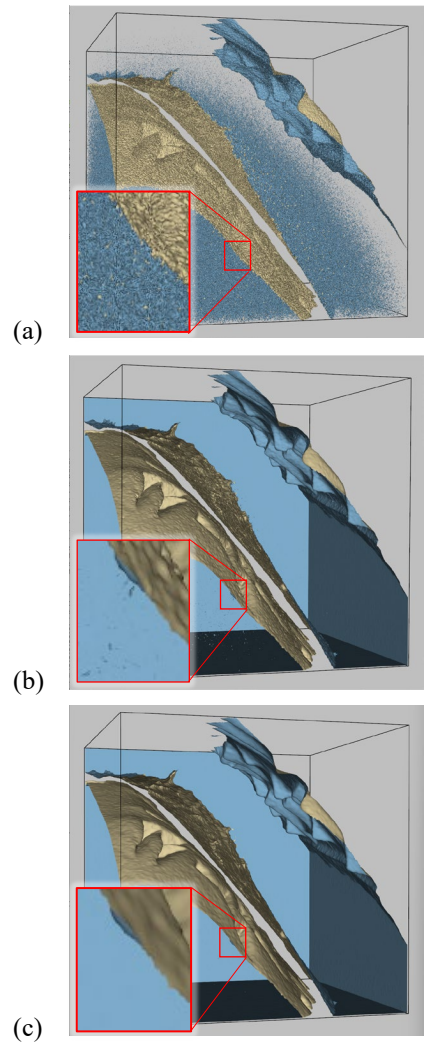

Figure 4: Isosurface of a cropped section of a preliminary 3D CT data set (512x512x512 from 1607x1950x2301 voxel, at offset (512, 512, 512), resulting in 65,835,427 triangles (a)), isosurface of Gauss-filtered and 2:1 resampled, same section (256x256x256, resulting 1,165,482 triangles (b)), and isosurface of the same Gauss-filtered section, denoised by iterated, joint bilateral filtering (result: 1,002,374 triangles (c)), all thresholding at the same scalar value.

## 3 IMPLEMENTATION

The workflow of processing of 3D volume, geometric, and feature data consists of the following steps:

1. Import (HDF5) and preprocessing (cropping, low-pass filtering, resampling, and denosing) of 3D volume data
2. Creation of an octree-based, compressed representation of isosurface and per-vertex data (we call it EXA file format)
3. Triangulation of the isosurface, implementing a 2-manifold dual marching cube algorithm (creating a mesh)
4. Postprocessing of the triangle mesh (smoothing, feature extraction, segmentation, and ambient occlusion)
5. Interactive 3D visualization of the segmented triangle mesh and per-vertex feature data (with optional stereoscopic rendering, head-tracking, gesture control)

All methods are implemented in C, taking advantage of multi-core parallelization in OpenMP and autovectorization. We utitlize OpenGL 4.6 for 3D rendering purposes, implementing fragment, vertex, and geometry shaders. GTK 3 is used for abstraction of the user interface. The modules are tested in Linux (Ubuntu/gcc) and Windows environments (MSYS2/gcc and MS Visual Studio).

### 3.1 Import and preprocessing of 3D volume data

The raw volume data, as result of CT, previously reconstructed on a regular 3D grid, is preprocessed, using methods of 3D image processing. Due to our decision to apply surface-orientated segmentation, it is more important to provide good course of the signal instead of remaining full resolution, which would be necessary for volume-oriented segmentation. Since our surface reconstruction stage achieves sub-voxel precision based on continuous signals, we consider low-passed filtering and 2:1 sampling (in all 3 dimension) the 3D volume data. This first step leads to 8:1 reduction of the volume. In the next step we perform edge-preserving denoising. We implemented several algorithms, based on median, bilateral, and joint bilateral filtering, which were tested. We also provide optionally iterative execution of bilateral filtering, since it is well-known that converges to a piecewise constant signal, which is advantageous in our use case.

Our tools support flexible configurations of the filter kernels (window size and functions) and its parameters. To exploit multi-level parallelism of modern workstation architectures, we tuned our software for efficient use of multi-threading (OpenMP) and vectorization (SIMD). To avoid conflicts in caching, and to enable autovectorization (gcc), we pay attention to memory locality and linearity of the data to be processed in the computationally intensive kernel operations.

Furthermore, we compared alternative filter kernels regarding computational effort and quality of the results. By replacing the usually applied Gaussian function for range weighting in the bilateral filtering by the Tukey function [10], we reduced the run time significantly, while the differences of the results – regarding the surface reconstruction – are negligible. We scaled the Tukey in relation to the Gaussian function, presuming to achieve the same attenuation for a σ' value in the Tukey as for the σ value in the Gaussian weighting (Figure 5).

This leads to

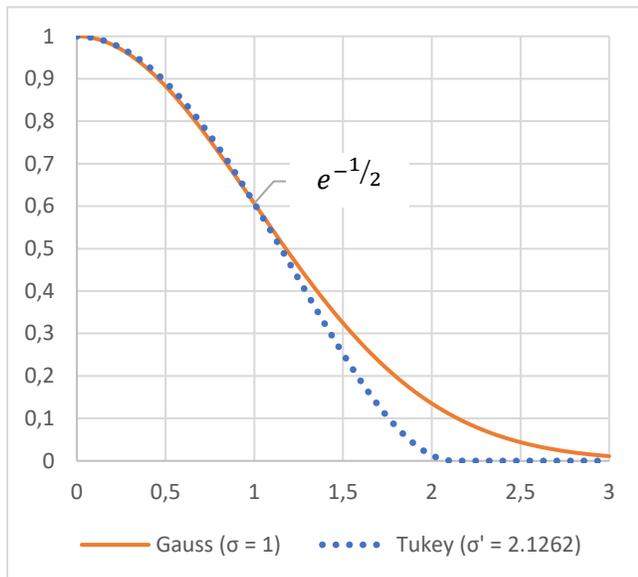

Figure 5: Gaussian vs. Tukey range weighting function.

$$Tukey(x) = \left(1 - \left(\frac{x}{\sigma'}\right)^2\right)^2 \text{ for } x < \sigma'; \text{ otherwise } 0$$

with $\sigma' = \frac{1}{\sqrt{1-\sqrt{e^{-\frac{1}{2}}}}} = 2.1262$

where $x$ is the difference between the image value at the currently focused location and the image value at the individual location of the considered window of the filter kernel.

By exploitation multi-level parallelism, memory locality and caching, as well as selection of effective, computationally less expensive methods, we achieved a speed-up of two orders of magnitude, compared with naïve implementation, which enables applications on mobile workstations equipped with modern processors.

Essential for successful automation of the processing pipeline is an appropriately estimated value of σ. In our approach, we assume Gaussian noise characteristics of the measured data. We analyzed its histogram curve, where typically two local maxima can be observed for our type of materials. After each of the filtering steps, the local maxima of the probability density are raised, and the standard deviation (= noise) is reduced (Figure 6).

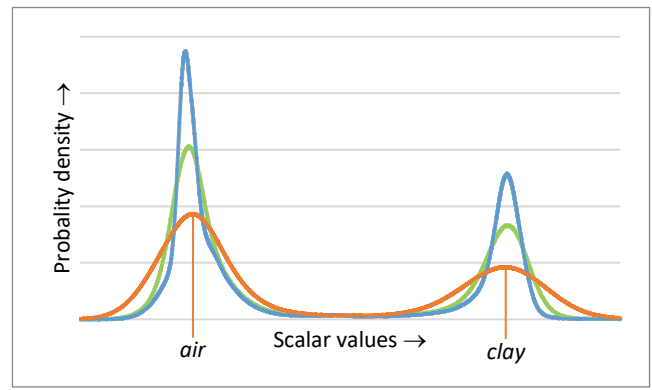

Figure 6: Histograms of measured CT data from an enclosed cuneiform clay tablet: (a) original values (orange), (b) low-pass filtered and 2:1 resampled (green), and (c) denoised by joint bilateral filtering (blue).

To estimate the σ value, we calculate the difference of the location of the upper maximum (representing clay material) and the location were the probability density falls to $e^{-1/2}$ of the probability density of the upper local maximum, according to the Gauss function. We decided to analyze the slope towards the local minimum (Figure 7).

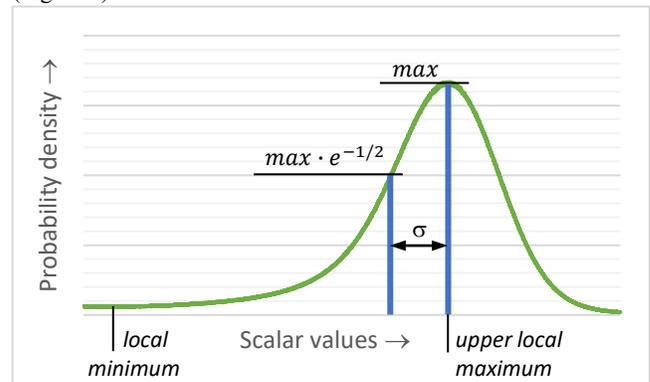

Figure 7: Estimation of the standard deviation σ, assuming Gaussian noise (zoomed cutoff of histogram, see Figure 6).

## 3.2 Compressed representation of isosurface

For extraction of the clay surface, we implemented an advanced isosurface algorithm. The aim of this method is to create a contouring representation in 3D. This results in reduction of the complexity from $O(n^3)$ to $O(n^2)$, where $n$ is the resolution of the volume data in each dimension. Successful application of this approach depends on the appropriate parameterization of the iso-threshold value $\tau$. To further automate the processing pipeline, we developed an estimator for $\tau$, which is optimized for our use case, where clay tablets are surrounded by air. The clay material itself may embed small areas of air or other material – e.g. organic – with less density than clay. However, $\tau$ should especially resolve the interspace between the tablet and the envelope.

Instead of implementing straightforward methods, such as using the mean of the values of the two peaks in the histogram, or adapting methods of image segmentation for our purposes, we developed a special method. In our experiments, we discovered that threshold values at the slope of the upper peak towards the local minimum of the histogram show good results. To estimate an appropriate threshold value, one can use the mean value where clay exists (= upper local maximum of probability density $pd$), minus $f \cdot \sigma$, with $f$ as a given constant (e.g. $f = 3$). In our case study, we implemented another approach: we identify the location in the histogram, where the $pd$ just exceeds its local minimum, multiplied by a given factor $f$ (e.g. $f = 2$, Figure 8).

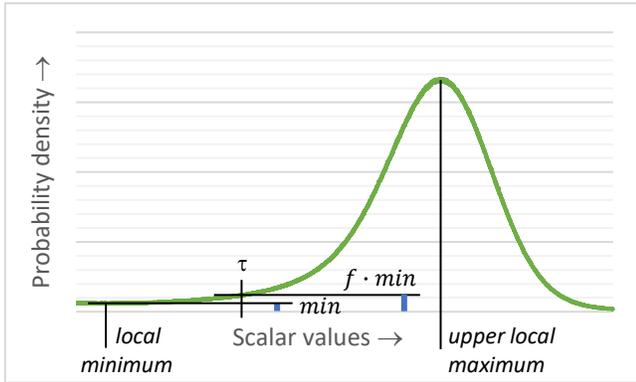

Figure 8: Estimation of the thresholding value $\tau$, looking for the location at the left slope of the upper peak where the probability density approaches its local minimum, multiplied by a given factor f (zoomed cutoff of histogram, see Figure 6).

As common in most isosurface algorithms, signs at each point of the regular 3D grid are analyzed, indicating if the respective scalar value at its position is less or greater than $\tau$. Active edges are defined where the sign changes along the edge. By considering similar aspects of parallelization as in section 3.1, we developed an efficient implementation. Our implementation avoids incremental allocation of memory by *malloc()*, e.g. where active cells etc. are identified. We replace this strategy by two passes. First, we count the number of required data elements and allocate sufficient memory afterwards. Second, we create the content of the elements on data locations we get by atomic capture operations. With the parallel implementation of theses steps, we get a speed-up of up to two orders of magnitude.

Inspired by [19], we integrate an efficient isosurface extraction algorithm, which prepares an octree-based representation, and finally creates an intermediate file in a format, which we call EXA. In contrast to [19], where sequential, context-based arithmetic compression is used, our demand was to develop parallel algorithms. Due to this requirement, we developed a static code table, using the 8-bit sign pattern of the cube vertices at the next course level of the octree at the respective location as context. This results in a transfer table to code the 8-bit signs, consisting of 65,536 entries. Each of them provides the number in a set of possible 8-bit signs which most likely can be predicted, resulting in a variable-length code. Moreover, only one octet of potentially eight children must be coded at each level of the octree, since all other seven sign patterns can be reconstructed from its neighborhood (Figure 9).

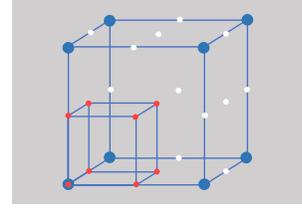

Figure 9: Signs at 8 grid points of a cell or octree cube (red) are encoded, based on a code table considering 8 context signs (blue) at the parental level of the octree hierarchy. 11 further grid points are reconstructed from the neighborhood (white).

Based on our precalculated code table, multi-level compression is not only parallelized by parallel prefix-sum scanning for calculation of the individual bit position, and afterwards atomic code writing. Furthermore, we get topology compression ratios similar to those mentioned in [19] (0.53 bit per vertex for a spherical volume data set on a $257^3$ grid). In Table 1, we show results, compressing an isosurface of a spherical scalar field. For example, in a 256x256x256 test case, we need approximately 0.87 bit per vertex, which means 0.435 bit per triangle in a mesh, without sub-voxel precision. Our code is complemented by 1 bit per ambiguous cell facets, which are resolved by "The asymptotic decider" [35]. Due to the rareness in practical use cases, its overhead is neglectable.

| Grid resolution (float values) | Number of vertices | Topology bits per vertex | Compression factor |
|---|---|---|---|
| 256x256x256 | 248,426 | 0.869 | 2,487 : 1 |
| 512x512x512 | 996,842 | 0.871 | 4,949 : 1 |
| 1024x1024x1024 | 3,994,274 | 0.866 | 9,936 : 1 |
| 2048x2048x2048 | 15,994,154 | 0.868 | 19,791 : 1 |

Table 1: Compressed isosurface of a spherical field (values: radius from center of volume, threshold: 90 % of max. radius).

To add geometric information, we quantize the normalized locations where the edges of the cells cross the isosurface, which results in approximately $N$ bits at each "active" edge, where $N$ is the precision in bit (max. normalized error: $0.5 \cdot 2^{-N}$). For a triangle mesh, this results in approximately $N/2$ bit per triangle. Due to the constant amount of precision data per edge, parallel coding is trivial.

By default, we use 8 bits. It makes sense to adapt this precision to the characteristics of the data set analyzed, e.g. noise and overall resolution. In our case study, where we process CT data from ancient cuneiform tablets, it turns out that a precision of 4 bits is sufficient, representing 16 equidistant steps along each active edge.

Our code is complemented by compact delta representations of vertex positions and normal vectors, which is used to store results of surface smoothing. We apply some sort of predictive coding, taking advantage of a variant of *Simple8b* for parallel variable-length encoding, using 64-bit and 128-bit words for delta compression of positions and normal, respectively. Each of them is filled with codes of constant length, specified in a selector [2]. Additional per-vertex data – e.g. for logarithmically quantized curvatures and classified shapes, partition numbers from segmentation, and precalculated ambient occlusion – are typically coded in 16-bit words: 7 bits for shape characteristics (14 curvature classes, multiplied by 9 shape classes, plus 1 for "flat"), 3 bits for up to 8 partition classes, and 6 bits for discretized ambient occlusion.

## 3.3 Triangulation of the isosurface

To enable the application of mesh smoothing algorithms and to create watertight surfaces for 3D print in STL file format, we aim to create triangle meshes with 2-manifold characteristics. In principle, isosurfaces can fulfill this prerequisite. But in detail, it is not straightforward to avoid violations. We develop a novel, table-based 2-manifold dual marching cubes algorithm, based on the isosurface data extraction described before. Our algorithm produces a quadrilateral around each active cell edge, as in [11]. As in previously developed dual marching cubes algorithms, one vertex is usually used for each of the four cells that are incident on the active edge, e.g. by calculating the centroid of all edge crossings connected to this part of the isosurface. To avoid non-manifold situations, at the respective cells we create two dual vertices. This is consequently done where ambiguous cell facets are detected while walking around the cell, connecting the edge crossings. This configuration occurs where all 4 edges are crossed (Figure 10) without the need to consider the neighborhood.

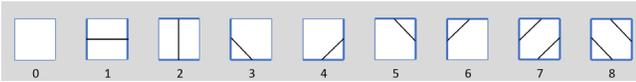

Figure 10: 9 cell facet configurations (0 = empty cell, 1-6 = unambiguous, 7-8 = ambiguous). Blue lines show active edges, where the sign changes between the vertices and the isosurface crosses. The black lines connect two crossings at active edges.

This leads to an extended quadrilateral (we call it "x-quad"), where each of the four vertices is potentially splitted into two, resulting in 4- to 8-sided polygons, which are represented by 2 to 6 triangles.

We precalculated a table consisting of all cell configurations, to accelerate our mesh generation and triangulation algorithms. By considering all combinations of cell facets, we identified 328 possible cases, which can be reduced to 27 basic cell configurations by omitting mirrored, rotated, or inverted variants. In Figure 11, all visualized x-quad faces are clipped to its respective quarter parts inside the cell. The configurations are sorted by number of 1 to 4 positive (or negative) cell vertices (5, …, 8 are equivalent to 3, …, 0), and by the number of faces crossing the cell. Its maximum is 4 (configuration 4N). The maximum number of dual vertices inside a cell is 4 (configuration 4J, with 2 faces, both consisting of 2 dual vertices). In several cell configurations (2B, 3B, 3C, 3E, 4D, 4E, 4F, and 4G, 4J, 4M), two dual vertices are created for a respective face, to avoid situations where non-manifold meshes could be produced. Even in cases were the neighborhood does not require it, this approach also leads to better reproduction of features in these cases, avoiding sharp crease angles. Depending on the respective active edge, one or two dual vertices are used for the mesh as part of the x-quad which is produced by our algorithm around the active edge. Even in the case of three dual vertices for one face (4E), no more than two dual vertices are necessary to represent the quarter part of each of the x-quads in this cell configuration.

In the initial publication of the marching cube algorithm, 15 basic cases were presented [25]. As later published, these can be reduced to 14, since one case can be removed by reflectional symmetry. The ambiguity of several of these cases can lead to cracks at ambiguous cell facets. This issue and the topologically correct construction of isosurfaces in general have been addressed in [7][9][21][24][34], resulting in 33 fundamental configurations.

Figure 10 illustrates our approach by walking around the contouring lines around the cell facets. This results in cyclic paths surrounding each of the up to 4 faces shown in Figure 11. Surface elements representing a tunnel-shaped topology inside the cell cannot be identified by this algorithm. Using the notation in [7], this regards to the cases 4.1.2, 6.1.2, 7.4.2, 10.1.2, 12.1.2, and 13.5.2.

Subtracting these six cases from the 31 basic cases (two cases from "Marching Cubes 33" can be removed by reflection symmetry), results in 25 cases. We identified 27 basic cases, i.e. two more, which had not been published at the time of implementing our algorithm. Meanwhile, these two additional cases (4F and 4J in Figure 11) were presented in [6] (cases 13.7 and 13.6.1 in their paper), and also two additional cases containing tunnels (cases 13.3.2 and 13.6.2 in their paper).

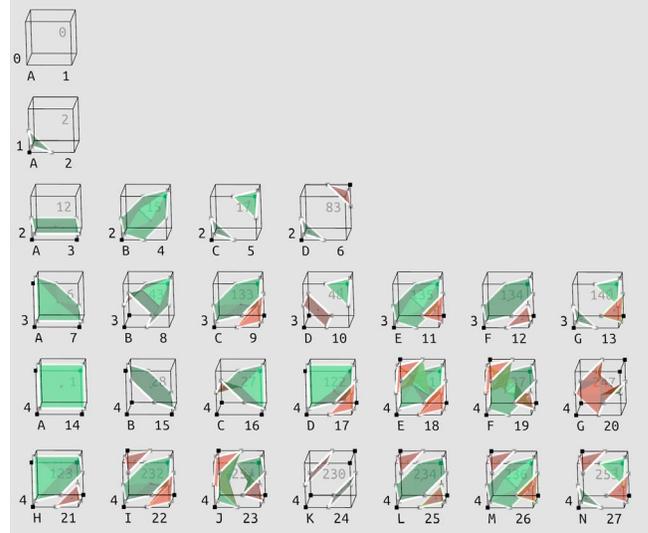

Figure 11: All 27 basic cell configurations we identified.

In our triangulation, we consider the two possible cases of crease orientations at each x-quad. We decide to use the most similar case, compared to the curvature, estimated from normal vectors at single dual vertices or using averaged positions and normal, where two dual vertices represent a cell. Of course, this applies only for convex quads. In case of a concave quadrilateral, we do not have an alternative.

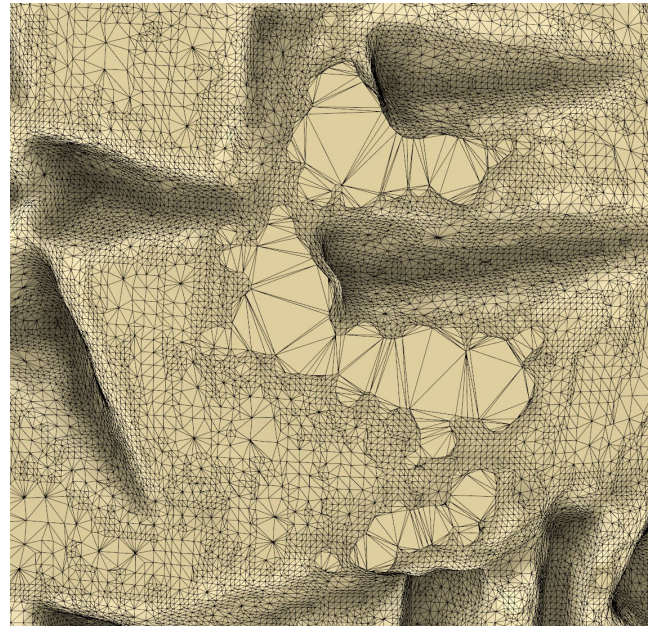

Figure 12: Result of our algorithm filling holes at segmentation boundaries, in combination with our adaptive vertex clustering algorithms, to reduce the polygon count at locations with less curvature.

Clockwise or counterclockwise ordering is chosen, depending on the gradient sign at the active edge where the respective x-quad is created. This guarantees consistent orientation throughout the mesh.

Neighborhood information is created based on the configuration table information on-the-fly, e.g. to link vertices to incident faces, which is required for further processing, such as interpolation of per-vertex normal vectors from per-face normal vectors at x-quads.

Finally, we implemented an option for adaptive vertex clustering, considering angular and positional deviation criteria, to reduce the number of polygons. Furthermore, we implemented a method based on the algorithm in [4][23], to fill holes of the surface at boundaries, which could caused by our surface-oriented segmentation (Figure 12). These additional triangles can be generated optionally.

## 3.4 Postprocessing of the surface mesh

### 3.4.1 Feature-preserving smoothing

For smoothing the surfaces, while preserving the features which are relevant in this application (here: high curvatures at the wedges of the cuneiform symbols), we implemented an alternative approach compared to existing methods. Instead of processing a triangular mesh, our algorithm is based on a mesh of x-quad faces, which are described in the previous section. We have tested a variety of iterative, median, and bilateral filter kernels to smooth the face-oriented normal vectors, assumed at the faces' center. Like in feature-preserving image filtering, the neighbored faces are considered.

Afterwards, the vertex positions are iteratively displaced, so that the angle between the line from the face center to the respective vertex position and the previously calculated face-oriented normal vector converges to 90°.

Finally, the per-vertex normal vectors are calculated as a weighted mean of the normal vectors of all faces which are incident to the respective vertex.

One exemplary result of our method is shown in Figure 13.

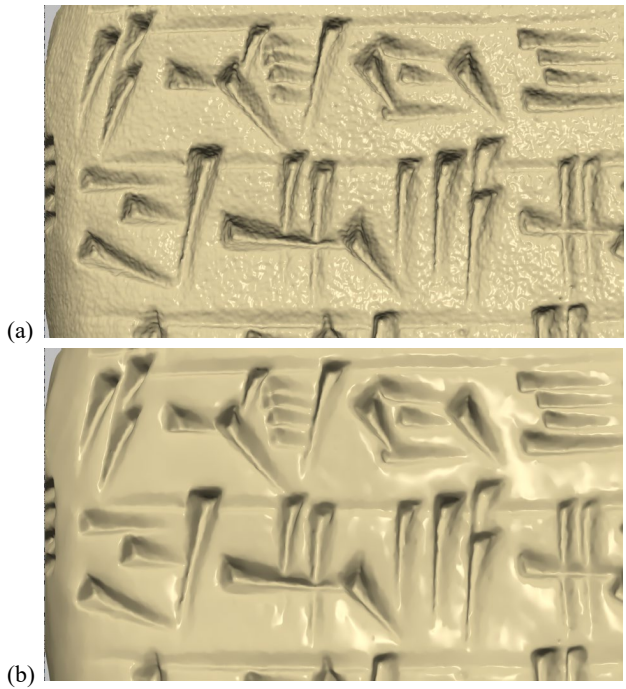

Figure 13: Visualization of isosurface of denoised 3D CT data (a) and smoothed isosurface mesh (b), using iterative weighted median filtering of face normals, and iterative vertex updating, rendered with a ambient/diffuse/specular lighting model shader in OpenGL.

### 3.4.2 Feature extraction

We decided that the most important feature to extract is the shape characteristic of the surface. Based on the results of mesh smoothing, we estimate the principal curvatures $k_1$ and $k_2$ at each vertex.

We apply a simple calculation of curvature between two vertices

$$curvature = \frac{(n_2 - n_1) \cdot (p_2 - p_1)}{|p_2 - p_1|^2}$$

We look for minimum and maximum values by using positions and normal vectors at pairwise opposite positions of the respective x-quad. At locations where a vertex at the x-quad is doubled, the mean values of the respective positions and normal vectors are used. The resulting face-oriented estimated principal curvatures $k_1$ and $k_2$ are used to create per-vertex curvature characteristics, by calculating the median of the principal curvatures $k_1$ and $k_2$ at its adjacent face, i.e. which are incident to the respective vertex, applying the iterative method presented in [50].

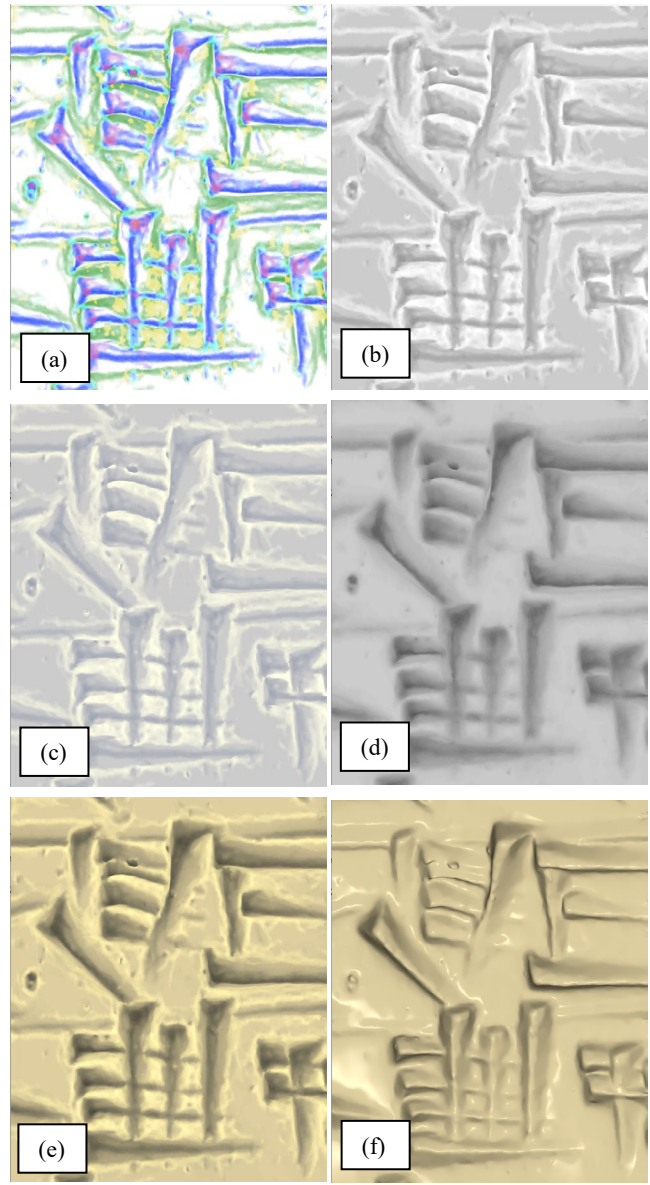

Figure 14: Illustration of curvature-oriented coloring (a) and darkening and highlighting (black-white (b) vs sky-blue shadows/sun-orange lights (c)), ambient occlusion (d), combination of both (e), and application of a simple ambient/diffuse/specular lighting model (f).

The two principal curvatures can be used to characterize the shape. Our method of coloring or emphasizing the cuneiform symbols by darkening and highlighting the shape characteristics is inspired by [17][18], where the shapes are classified according to
$$\varphi = atan2(k_1, k_2).$$
We linearly discretize the range of possible values of $\varphi$ into 9 categories. The most interesting cases of surface shapes are illustrated in Figure 15.

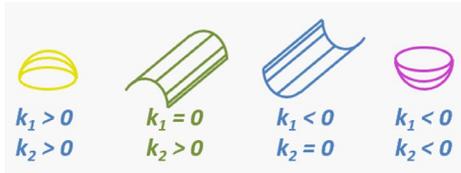

Figure 15: Main cases of convex and concave surface shapes.

Additionally, we calculate the curvedness value (total curvature)
$$c = \sqrt{k_1^2 + k_2^2},$$
which is discretized using 14 logarithmically scaled steps, each representing a relative increase of $\sqrt{2}$. We found out, that in our use cases, the range of 6.5 octaves and steps of ½ octave are sufficient. In sum, this results in $9 \cdot 14 + 1 = 127$ categories (one for "flat").

As illustrated in Figure 14, we use this 7-bit attribute for (a) per-vertex color-mapping, and for darkening and high-lighting in (b) black/white or (c) sky-blue/sun-orange to emphasize the cuneiform symbols. The latter is inspired by the paintings of the French artist Édouard Manet (1832–1883). The result appears similar to radiance scaling [13][56], but by using the precalculated 7-bit attribute, a much simpler table-based mapping shader can be used, and we are able to export precalculated colors as part of open file formats (e.g. PLY) to be visualized or processed in other tools or platforms (e.g. Sketchfab, which was tested in [37]).

### 3.4.3 Segmentation

Usually, the enclosed cuneiform tablet and the envelope are represented in one connected mesh at locations where they are touching or glued. To split the mesh into separate partitions, e.g. to visualize these objects separately, an appropriate segmentation method is required.

Our segmentation algorithm is based on a curvature threshold, e.g. we apply a negative threshold value of the principal curvature $k_1$, representing the magnitude of the concavity, to identify the boundaries. As such, the extracted partitions of the surface can contain holes, e.g. where the gap between the tablet and the envelope cannot be resolved, or no gap exists at all (touched or glued).

### 3.4.4 Ambient occlusion

One further precalculated per-vertex attribute represents occlusion of surrounding light. The usage of a discretized version of the value is incorporated in the rendering process. It can be mapped directly to the per-vertex intensity, as in Figure 14 (d), or combined with other table-based mapping methods, as in Figure 14 (e). Both examples do not require any lighting calculation, in contrast to (f), where a Phong lighting models is implemented for comparison.
Taking advantage of our octree-based representation, we implemented a parallel ambient occlusion analyzer, based on [28][29]. Several rays, equidistantly distributed over a hemisphere are analyzed for crossing of another surface inside a given radius. In our algorithm, we create the orientations of a given number of rays according to a Fibonacci spiral method, to achieve uniform distribution of direction on the hemisphere [27].

As part of the integration process, a weighting function considers the angle between the individual ray, which is not occluded, and the normal vector at the considered vertex ($\varphi$).

For our use case, we developed a special weighting function, which increased the recognition of shadows significantly. Instead of Lambert's cosine law commonly implemented as
$$w_1(\varphi) = cos(\varphi),$$
we use this function:
$$w_1(\varphi) = sin(\varphi) \cdot \sqrt{cos(\varphi)}.$$
Both functions are illustrated in Figure 16.

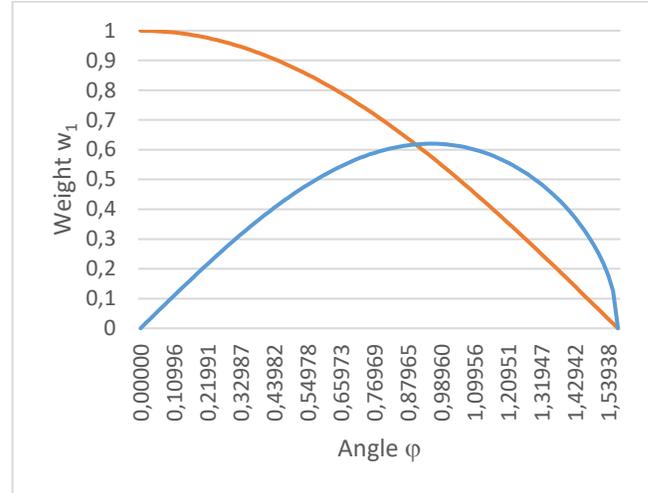

Figure 16: Weighting function to increase recognition of shadowing (blue), in comparison with Lambert's cosine law (orange).

Its result is multiplied by a linearly decreasing value, which depends on the distance $d$ of the nearest surface, crossed by the respective ray. This weighting function becomes 0 at a given maximal radius $r$:
$$w_2(d) = 1 - d/r \text{ for } d < r, \text{ and otherwise: } w_2(d) = 0$$

## 3.5 Interactive 3D visualization

We developed an interactive 3D visualization application, specialized for our use case. It supports real-time rendering of surface representations, which can be imported in our proprietary file format EXA, or in open file formats, such as PLY. The triangulated surface is rendered by using geometry, vertex, and fragment shaders, taking advantage of several features of OpenGL 4.6. For interactive navigation of the 3D model (rotation, translation, zoom), mouse control is used in conjunction with shift and control keys. Numerous menus and dialogs have been implemented in GTK3, which abstracts the underlying platform (operating system, window manager, etc.). They allow to parameterize the shader functions, e.g. how the shape features should be visualized (color-table mapping, contrast), to configure the lighting model and further rendering options (e.g. culling, clipping, fog, transparency), and creation of animations. The partitions, as result of our segmentation, can be visualized separately, switching them on or off, and assigning separate colors.

We also support stereoscopic displays, either by using OpenGL quad-buffering (e.g. for zSpace AIO) or by incorporating interfaces for special devices. One example is the integration of an SDK to support head-tracking autostereoscopic displays, such as the SpatialLabs products from Acer (15,6" 4K 3D laptops and displays) and the SR Pro² from Leia (32" 8K 3D display). These displays support head-tracked, glass-less stereoscopic viewing, based on the SR technology by Dimenco/Leia. Additionally, we support the Looking Glass display, which proves some sort of light field technology. Furthermore, we integrated gesture control, supporting the LeapMotion hand tracking devices from UltraLeap.

## 4 RESULTS

The development and construction of a portable high-resolution X-ray micro-CT scanner allowed us to transport and set up the scanner at the Louvre Museum and scan a dozen ancient enveloped clay tablets. Figure 17 shows one of the scans selected to present the following results. Our visualization of the micro-CT data allowed the Assyriologist to read the hidden text on the tablet, rotate the object on a 3D screen to decipher all of its sides (Figure 18), and even read the written artifact while holding its 3D print as if it were the original tablet (Figure 19). The resolution of the details is very fine, allowing for palaeographic analysis. In addition, the rendering of the seal impressions on the envelopes reveals details that are not always apparent to the naked eye. We provide a video clip in [39].

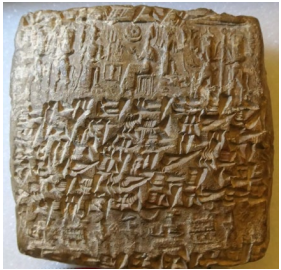

Figure 17: Envelope of a loan contract, Central Anatolia, 19th century BC (Musée du Louvre, photo: Cécile Michel).

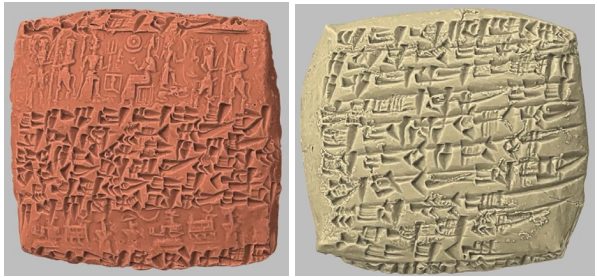

Figure 18: Visualization of the segmented envelope (12.6 mio. triangles) and hidden tablet (3.2 mio. triangles; sum: 20.0 mio. tri.).

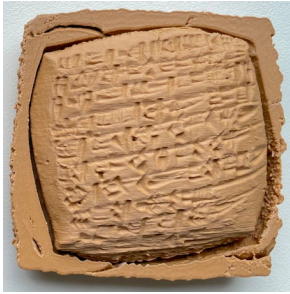

Figure 19: 3D print of the tablet and its envelope.

To evaluate the performance of processing data acquired by X-ray CT to prepare interactive visual exploration, e.g. virtual unpacking of an enclosed cuneiform tablet, we use an HDF5 file representing the tomographic reconstruction of the artifact AO8295 (Figure 17, 5.9 x 6.1 x 2.1 cm) on a regular 3D grid (resolution: 38 μm, 12.9 GB) [32]. Table 2 summarizes the run-time analysis performed on an Acer SpatialLabs ConceptD CN715-73G mobile workstation, consisting of an Intel i7-11800H CPU (8 cores, 16 threads), Nvidia GeForce RTX 3080 GPU, 64 GB memory, and 2 TB SSD (Samsung 980 Pro), using Windows 11 Pro 23H2. Our software is compiled with gcc 14.2.0, and run under MSYS2, using 16 threads (OpenMP). The sources are publicly available in [40].

|  | Resolution / Volume | Run time |
|---|---|---|
| Import of HDF5 file | 1871 x 999 x 1723 | 6.699 sec |
| Low-pass filtering (Gauss, 3x3x3 kernel) and resampling (2:1) | → 935 x 499 x 861 (SNR: 17 dB → 23 dB) | 0.601 sec |
| Estimation of signal to noise ratio | 935 x 499 x 861 | 0.233 sec |
| Cross-bilateral denoising (Tukey, 3x3x3 kernel, joint image: Gauss, 3x3x3 kernel), 2 iterations | 935 x 499 x 861 (SNR: 23 dB → 28 dB) | 5.757 sec |
| Estimation of threshold value | 935 x 499 x 861 | 0.232 sec |
| Isosurface extraction (threshold: 0.000839996) | 935 x 499 x 861 → 10,011,300 vertices | 1.273 sec |
| Export of EXA file (4 bit precision of normalized positions of edge crossings) | 6.448 MB, 1.168 bit/v. topology + 3.984 bit/v. precision | 0.696 sec |
| Import of EXA file | 6.448 MB, 5.152 bit/v. | 1.277 sec |
| Initial creation of vertex positions and normal vectors | 10,011,300 vertices | 1.770 sec |
| Smoothing of face normals, using weighted median filtering, 32 iterations | 10,011,300 vertices | 9.636 sec |
| Calculation of vertex normals | 10,011,300 vertices | 0.224 sec |
| Updating vertex positions (8 it.) | 10,011,300 vertices | 2.663 sec |
| Calculation of vertex curvatures | 10,011,300 vertices | 0.425 sec |
| Mesh traversal, including segmentation ($k_1 < -0.5$) | 10,011,300 vertices | 1.639 sec |
| Export of EXA file, including positional deltas (19.9 b/v), normal deltas (19.1 b/v), and additional features (16.1 b/v) | 754 MB 60.223 bit/vertex | 1.793 sec |
| Import of EXA file | 754 MB, 60.223 bit/v. | 1.476 sec |
| Initial creation of vertex positions and normal vectors | 10,011,300 vertices | 0.927 sec |
| Decompression of vertex positions and normal vectors | 10,011,300 vertices | 0.183 sec |
| Analysis of *ambient occlusion* (160 rays per vertex, max. distance: 64 grid steps) | 10,011,300 vertices | 28.6 min |
| Export of EXA file, with added ambient occlusion values | 759 MB 60.671 bit/vertex | 2.058 sec |

Table 2: Run times of the main processing steps (Sec. 3.1 to 3.4).

## 5 CONCLUSION

The methods we have developed and efficiently implemented, exploiting the multi-level parallel performance and cache architecture provided by modern workstations, have proven to be useful for processing micro-CT datasets at collection sites. Our visualization component enables interactive 3D rendering, supporting virtual unpacking of enclosed cuneiform tablets and interdisciplinary understanding. For intuitive usage, we support immersive techniques, such as autostereoscopic displays and head/hand/finger tracking.

## ACKNOWLEDGMENTS

The project "Reading Closed Cuneiform Tablets Using High-Resolution Computed Tomography" (RFA09) was funded by the Deutsche Forschungsgemeinschaft (DFG, German Research Foundation) under Germany´s Excellence Strategy – EXC 2176 'Understanding Written Artefacts: Material, Interaction and Transmission in Manuscript Cultures', project no. 390893796. The research was conducted within the scope of the Centre for the Study of Manuscript Cultures (CSMC) at Universität Hamburg. Many thanks to Ariane Thomas, head of the Département des Antiquités Orientales at the Musée du Louvre, and Véronique Pataï who has assisted us during the campaign in February 2024.

## REFERENCES

[1] Anderson, E., S. and Levoy, M.: *Unwrapping and Visualizing Cuneiform Tablets*. IEEE Computer Graphics and Applications, November/December 2002.

[2] Anh, V. N. and Moffat, A.: *Index Compression Using 64-Bit Words*. Software: Practice and Experience, 40(2):131–147, 2010.

[3] Avizo Software. https://www.thermofisher.com/de/de/home/electron-microscopy/products/software-em-3d-vis/avizo-software.html, accessed 16.06.2024.

[4] Barequet, G. and Sharir, M.: *Filling Gaps in the Boundary of a Polyhedron*. Computer-Aided Geometric Design, 12(2):207-229, March 1995.

[5] Béranger, M.: *Fonctions et usages des enveloppes de lettres dans la Mésopotamie des IIIe et IIe mil. Av. J.-C. (2340-1595 av. J.-C.)*. Épistolaire 44, Librairie Champion, Paris, pp. 25-43, 2018.

[6] Chen, Z. and Zhang, H.: *Neural Marching Cubes*. ACM Trans. on Graphics, Vol. 40, No. 6, 2021.

[7] Chernyaev, E. V.: *Marching Cubes 33: Construction of Topologically Correct Isosurfaces*. Technical Report CN/95-17, Institute of High Energy Physics, Russia, Presented at GRAPHICON'95, Saint-Petersburg, 1995.

[8] CuneiformAnalyzer. www.cuneiform.de, accessed 16.06.2024.

[9] Custodio, L., Pesco, S., and Silva, C.: *An extended triangulation to the Marching Cubes 33 algorithm*. Journal of the Brazilian Computer Society, 25:6, 2019.

[10] Durand, F. and Dorsey, J.: *Fast Bilateral Filtering for the Display of High-Dynamic-Range Images*. SIGGRAPH 2002.

[11] Grosso, R. and Zint, D.: *A parallel dual marching cubes approach to quad only surface reconstruction*. The Visual Computer 38, 1301-1316, 2022.

[12] Engel, K., Westermann, R., and Ertl, T.: *Isosurface Extraction Techniques for Web-based Volume Visualization*. Proc. IEEE Visualization, 1999.

[13] Fisseler, D., Müller, G., and Weichert, F.: *Web-Based Scientific Exploration and Analysis of 3D Scanned Cuneiform Datasets for Collaborative Research*. Informatics 4(4), December 2017.

[14] Fleishman, S., Drori, I., and Cohen-Or, D.: *Bilateral Mesh Denoising*. ACM Transactions on Graphics, 2003.

[15] GigaMesh. www.gigamesh.eu, accessed 16.06.2024.

[16] *GigaMesh Tutorial 09. Unpacking a Cuneiform Tablet*. https://www.youtube.com/watch?v=0jqP_6jyjyo, uploaded 26.07.2019, accessed 16.06.2024.

[17] Koenderink, J. J. and van Doorn, A. J.: *Surface Shape and Curvature Scales*. Image and Vision Computing, Vol. 10, Issue 8, October 1992.

[18] Kindlmann, G., Whitaker, R., Tasdizen, T., and Moller, T.: *Curvature-Based Transfer Functions for Direct Volume Rendering: Methods and Applications*. Proc. IEEE Visualization, 2003

[19] Lee, H., Desbrun, M., and Schröder, P.: *Progressive Encoding of Complex Isosurfaces*. Proc. ACM SIGGRAPH 2003.

[20] Lee, Y. and Wang, W.-P.: *Feature-preserving Mesh Denoising via Bilateral Normal Filtering*. Proc. IEEE International Conference on Computer-Aided Design and Computer Graphics, 2005.

[21] Lewiner, T., Lopes, H., Vieira, A., and Tavares, G.: *Efficient Implementation of Marching Cubes' Cases with Topological Guarantees*. Journal of Graphics Tools, 8:2, pp 1-15, 2003.

[22] Lewiner, T., Lopes, H., Velho, L., and Mello, V.: *Simplicial isosurface compression*. Proc. Vision, Modeling, and Visualization, 2004.

[23] Liepa, P.: *Filling Holes in Meshes*. Eurographics Symposium on Geometry Processing, 2003.

[24] Lopes, A. and Brodlie, K.: *Improving the Robustness and Accuracy of the Marching Cubes Algorithm for Isosurfacing*. IEEE Trans. on Visualization and Computer Graphics, Vol. 8, No. 1, 2003.

[25] Lorensen, W. E., Cline, H. E.: *Marching cubes: A high resolution 3D surface construction algorithm*. ACM SIGGRAPH Computer Graphics, Volume 21, Issue 4, 1987.

[26] Lu, X., Liu, X., Deng, Z., and Chen, W.: *An Efficient Approach for Feature-preserving Mesh Denoising*. Optics and Lasers in Engineering, Volume 90, pp. 186-195, 2017.

[27] Marques, R., Bouville, C., Bouatouch, K., and Blat, J.: *Extensible Spherical Fibonacci Grids*. IEEE Transactions on Visualization and Computer Graphics, Volume 27, Issue 4, 2021.

[28] Méndez-Feliu, A. and Sbert, M.: *Efficient rendering of light and camera animation for navigation a frame array*. Proc. Computer Animation and Social Agents (CASA 2006), 2006.

[29] Méndez-Feliu, A. and Sbert, M.: *From obscurances to ambient occlusion: A survey*. The Visual Computer, 25(2):181-196, 2009.

[30] Michel, C.: *Making Clay Envelopes in the Old Assyrian Period*. Integrative Approaches to the Archaeology and History of Kültepe-Kanesh, Kültepe, 4-7 August, 2017, Fikri Kulakoğlu, Cécile Michel, and Güzel Öztürk (eds), Brepols, Turnhout, pp. 187-203, 2020.

[31] Michel, C.: *Ecrire sur argile. La matérialité des textes cuneiforms*. Argiles. De la physique du matériau à l'expérimentation, Xavier Faivre (ed.), Archaeopress Archaeology, Oxford, pp. 98-111, 2023.

[32] Michel, C., Schroer, C., Olbrich, S., Ehteram, S., and Beckert, A.: *AO 8295 (X-Ray Tomography 3D data of an Enveloped Clay Tablet, Louvre Museum, Paris)*. Data set, 2024. http://doi.org/10.25592/uhhfdm.14776

[33] Ngan-Tillard, D.: *Reading 4000 years old clay tablet through intact envelope using X-ray micro-CT scans*. https://www.youtube.com/watch?v=qvoZQVw6VKs, uploaded 30.04.2018, accessed 16.06.2024.

[34] Nielson, G. M.: *On Marching Cubes*. IEEE Trans. on Visualization and Computer Graphics, 2003.

[35] Nielson, G. M., Hamann, B.: *The Asymptotic Decider: Resolving the Ambiguity in Marching Cubes*. Proc. IEEE Conf. on Visualization, 1991.

[36] NINO Leiden (2018): *Seeing through clay: 4000 year old tablets in hypermodern CT scanner*. https://www.nino-leiden.nl/message/seeing-through-clay-4000-year-old-tablets-in-hypermodern-ct-scanner, accessed 16.06.2024.

[37] Olbrich, S.: Reconstruction of enclosed cuneiform tablet. https://sketchfab.com/3d-models/reconstruction-of-enclosed-cuneiform-tablet-31b0b8e4073241ecbd82dc3764c1739b, uploaded 10.09.2022, accessed 20.06.2024.

[38] Olbrich, S., Michel, C., and Schroer, C.: *Non-invasive unpacking of enclosed cuneiform tablets by visualization and printing of extracted and segmented surfaces from 3D CT volume data*. DOT 2022 – Deutscher Orientalistentag, September 12 – 17, 2022, Berlin, Germany.

[39] Olbrich, S. and Beckert, A.: *Non-invasive, virtual unpacking of enclosed cuneiform – 3D reconstruction and visualization of clay tablets based on data acquired by portable micro-CT scanner*. Video animation, 2024. http://doi.org/10.25592/uhhfdm.14772

[40] Olbrich, S. and Beckert, A.: *EXAVIS42 – Efficient methods for creation, feature extraction, and interactive visualization of isosurfaces of 3D volume data*. Software sources, 2024. http://doi.org/10.25592/uhhfdm.14778

[41] Pacanowski, R., Granier, X., and Schlick, C.: *Radiance Scaling for Versatile Surface Enhancement*. Proc. Symposium on Interactive 3D Graphics, 2010.

[42] Paris, S., Kornprobst, P., Tublin, J., and Durand, F.: *Bilateral Filtering: Theory and Applications*. Computer Graphics and Vision, Vol. 4, No. 1, 2008.

[43] Poston, T., Wong, T.-T., and Heng, P.-A.: *Multiresolution Isosurface Extraction with Adaptive Skeleton Climbing*. Computer Graphics Forum, Vol. 17, No. 3, September 1998.

[44] Rolff, T., Rautenhaus, M., Olbrich, S., and Frintrop, S.: *Segmenting Computer-Tomographic Scans of Ancient Clay Artefacts for Visual Analysis of Cuneiform Inscriptions*. Proc. 25th International Symposium on Vision, Modeling, and Visualization, 2020.

[45] Saupe, D. and Kuska, J.-P.: *Compression of Isosurfaces for Structured Volumes*. Proc. Vision, Modeling, and Visualization Conf., 2001.

[46] Schaefer, S. and Warren, J.: *Dual Marching Cubes: Primal Contouring of Dual Grids*. Proc. Pacific Conference on Computer Graphics and Applications, 2004.

[47] Schaefer, S., Ju, T., and Warren, J.: *Manifold Dual Contouring*. IEEE Transactions on Visualization and Computer Graphics, Volume 13, Issue 3, 2007.

[48] Schroeder, W., Martin, K., and Lorensen, B.: *The Visualization Toolkit ($4^{th}$ ed.)*. Kitware, 2006.

[49] Seim, H., Kainmueller, D., Heller, M., Lamecker, H., Zachow, S., and Hege, H.-C.: *Automatic Segmentation of the Pelvic Bones from CT Data Based on a Statistical Shape Model.* Eurographics Workshop on Visual Computing for Biomedicine, 2008.

[50] Spence, C. and Fancourt, C.: *An Iterative Method for Vector Median Filtering*. IEEE International Conference on Image Processing, 2007.

[51] Sun, X., Rosin, P. L., Martin, R., and Langbein, R.: *Fast and Effective Feature-Preserving Mesh Denoising*. IEEE Transactions on Visualization and Computer Graphics, Vol. 13, Issue 5, pp 925-938, 2007.

[52] Taubin, G.: *BLIC: Bi-Level Isosurface Compression*. Proc. IEEE Visualization, 2002.

[53] Titschack, J., Baum, D., Matsuyama, K., Boos, K., Färber, C., Kahl, W.-A., Ehrig, K., Meinel, D., Soriano, C., and Stock, S. R.: *Ambient occlusion – A powerful algorithm to segment shell and skeletal intrapores in computed tomography data*. Computers & Geosciences, Vol. 115, June 2018.

[54] Tomasi, C. and Manduchi, R.: *Bilateral filtering for gray and color images*. Proc. IEEE Conf. on Computer Vision, 1998.

[55] TU Delft (2018): *Cuneiform in a scanner*. https://www.tudelft.nl/en/delft-outlook/articles/cuneiform-in-a-scanner, accessed 16.06.2024.

[56] Vergne, R., Pacanowski, R., Barla, P., Granier, X., and Schlick, C.: *Radiance Scaling for Versatile Surface Enhancement*. Proc. Symposium on Interactive 3D Graphics, 2010.

[57] Yagou, H., Belyaev, A. G., and Wie, D.: *Mesh Median Filter for Smoothing 3-D Polygonal Surfaces*. Proc. Conf. Cyber Worlds, 2002.

[58] Yagou, H., Ohtake, Y., and Belyaev, A. G.: *Mesh Smoothing via Mean and Median Filtering Applied to Face Normals*. Proc. Geometric Modeling and Processing (GMP 2002), 2002.

[59] Zheng, Y., Hongbo, F., Au, K.-C., and Tai, C.-L.: *Bilateral Normal Filtering for Mesh Denoising*. IEEE Transactions on Visualization and Computer Graphics, Vol. 17, Issue 10, 2010.